\documentclass[aps,pre,twocolumn,groupedaddress,nofootinbib]{revtex4}

\usepackage{graphicx}
\usepackage{graphics}
\usepackage{amsfonts}
\usepackage{amsmath}
\usepackage{amssymb}
\usepackage{color}
\usepackage{soul}

\setcitestyle{super}

\newcommand{\Qtot}{{Q}_{\rm tot}}

\newcommand*{\revadd}[1]{{{#1}}}

\begin{document}

\title{The Electric Double Layer Has a Life of Its Own}

\author{C\'eline Merlet$^{1,2,3}$}
\author{David Limmer$^4$}
\author{Mathieu Salanne$^{1,2}$}
\author{Ren\'e van Roij$^5$}
\author{Paul A. Madden$^6$}
\author{David Chandler$^7$}
\author{Benjamin Rotenberg$^{1,2,*}$}
\affiliation{$^1$ \small Sorbonne Universit\'es, UPMC Univ. Paris 06, CNRS, UMR 8234 PHENIX, 75005 Paris, France}
\affiliation{$^2$ \small R\'eseau sur le Stockage Electrochimique de l'Energie (RS2E), FR CNRS 3459, France}
\affiliation{$^3$ \small Department of Chemistry, University of Cambridge, Cambridge CB2 1EW, UK}
\affiliation{$^4$ \small Princeton Center for Theoretical Science, Princeton, NJ-08544, USA }
\affiliation{$^5$ \small Institute for Theoretical Physics, Utrecht University, 3584 CE Utrecht, The Netherlands }
\affiliation{$^6$ \small Department of Materials, University of Oxford, Oxford OX1 3PH, UK}
\affiliation{$^7$ \small Department of Chemistry, University of California, Berkeley, CA-94720, USA}
\affiliation{$^*$ \small Corresponding author: benjamin.rotenberg@upmc.fr}

\date{\today}

\begin{abstract}
Using molecular dynamics simulations with recently developed importance sampling
methods, we show that the differential capacitance of a model ionic liquid based
double-layer capacitor exhibits an anomalous dependence on the applied
electrical potential. Such behavior is qualitatively incompatible with standard
mean-field theories of the electrical double layer, but is consistent with
observations made in experiment. The anomalous response results from structural
changes induced in the interfacial region of the ionic liquid as it develops a
charge density to screen the charge induced on the electrode surface. These
structural changes are strongly influenced by the out-of-plane layering of the
electrolyte and are multifaceted, including an abrupt local ordering of the ions
adsorbed in the plane of the electrode surface, reorientation of molecular ions,
and the spontaneous exchange of ions between different layers of the electrolyte
close to the electrode surface. The local ordering exhibits signatures of a
first-order phase transition, which would indicate a  singular charge-density
transition in a macroscopic limit.\\ 

\end{abstract}


\maketitle

\section{Introduction}
The electric double layer is generally viewed as simply the boundary that
interpolates between an electrolyte solution and a metal surface. In contrast,
we show here that the interface between an electrode and a model ionic liquid
exhibits behaviors distinct from that of the bulk liquid.  At a particular
voltage, the double layer undergoes an abrupt local ordering transition.  The
emergent ordered structure is consistent with that of room temperature ionic
liquids found on mica surfaces~\cite{liu_coexistence_2006}.  At another voltage,
the interfacial charge density changes abruptly.  This change appears to be
consistent with observations of anomalous electrical capacitance observed
experimentally~\cite{su_double_2009}.  Further, we show that this anomaly grows
with increasing electrode surface area, suggesting a singular charge-density
transition in a macroscopic limit.  These nonlinear electrochemical responses,
whose molecular origins we elucidate, offer a view of the electric double layer
that deviates significantly from standard mean-field models.

Generally, when an electrical potential is applied to an electrode in contact with an
electrolyte it induces a local, partial demixing of the cations and anions at
the interface. The surface charge is equal and opposite to the net charge in the
double layer and thus at a given potential reflects the facility of the
interfacial fluid to adopt a structure giving rise to that charge. 
The canonical model for the interfacial region of an electrolyte
at an electrode surface, the electric double layer, is that of Gouy and Chapman
(GC)~\cite{parsons_review-doublelayer_1990}. This mean-field model applies to a
dilute solution of point ions and predicts that the degree of demixing induced
by the voltage decreases exponentially with distance from a planar electrode
surface and that the differential capacitance of the electrode-electrolyte
double-layer increases monotonically with the magnitude of the applied electrode
potential.  Recently, there has been a great deal of interest in electrochemical
interfaces of room temperature ionic liquids (RTILs) driven in part by the
potential for constructing supercapacitors that exploit this charge separation
to create high power energy storage devices~\cite{simon2008a}.  
Far from a collection of noninteracting point charges RTILs are dense mixtures
of ions, whose physical properties are largely determined by correlations
between ions.

Effects arising from correlations within RTILs at electrochemical interfaces have been
observed experimentally and in simulation~\cite{merlet2013c,fedorov_ionic_2014}. 
Specifically, layering has been observed in X-ray reflectivity
experiments~\cite{mezger2008a,uysal_structural_2014}, is detected in surface
force and AFM experiments~\cite{atkin2007a,perkin2010a,perkin2012a,smith2013a}
and has been examined in a number of computer simulation
studies~\cite{pinilla2007a,fedorov2008a,vatamanu_molecular_2010,mendonca2012a,padua2013a}.
Some of the latter have illustrated  how the polarization of the charge density
in an ionic liquid in response to an electrode potential is affected by this
layering and the strong local coulomb ordering of the ions~\cite{lanning2004a,
pinilla2007a}. 
\revadd{In-plane ordering of the ions in RTIL has been detected
at the liquid-vapor interface by X-ray scattering~\cite{jeon_surface_2012},
on mica surfaces by AFM~\cite{liu_coexistence_2006}, 
and on gold electrodes by STM~\cite{pan2006a}.
Some} simulations~\cite{pounds2009b,kislenko2009a,tazi2010a,jha2013a} have reproduced this effect
and indicated a connection to changes in the electrode charge density.
\revadd{Using in-situ STM and AFM measurements, Atkin, Endres and co-workers 
underlined the influence of phenomena such as surface reconstruction 
(\textit{e.g.} with gold electrodes) and of impurities on the interfacial
structure~\cite{atkin_situ_2011,endres_interface_2011} and demonstrated
the advantage of combining these techniques with electrochemical impedance
spectroscopy to investigate these aspects~\cite{druschler2012a}.}

Not surprisingly then, the experimentally observed dependence of the differential
capacitance on potential has often been found to depart markedly from the
predictions of GC theory, often decaying with increasing potential and
exhibiting sharp peaks that are interpreted as abrupt changes in the local ionic
structuring. 
Efforts to extend GC theory to include effects from packing of ions have been
made at the mean-field level with some success~\cite{fedorov_ionic_2014}. These account
approximately for layering at the interface~\cite{kornyshev2007a} and the
tendency for local coulombic ordering of the ions at high
voltage~\cite{lauw_room-temperature_2009,bazant2011a,skinner_capacitance_2010}.
Specifically, Kornyshev and
coworkers have demonstrated the effect of these extensions on the differential
capacitance, qualitatively reproducing some features observed in
RTILs~\revadd{\cite{kornyshev2007a,bazant2011a}}.
However, correlations in-plane and effects from fluctuations have not been
considered, omitting molecular detail and its role in determining the
electrochemical properties of these systems. Such features are however essential 
in conditions where the liquid undergoes large structural
reorganization, which as we discuss below, are intimately connected to
non-monotonic features of the differential capacitance.  

Here we address the correspondence between structure, fluctuations and
electrochemical response in RTILs directly by applying novel simulation methods. Such
methods are robust and free of assumptions related to the relative importance of
correlations that constrain mean-field models. In
section~\ref{sec:systemandmethods} we first describe the considered system and
present the theoretical approach to determine the evolution of observables, such
as the differential capacitance or structural properties, with applied voltage.
Section~\ref{sec:results} then reports the main results of this study.

\section{Methods}
\label{sec:systemandmethods}

\begin{figure*}[ht!]
\begin{center}
\includegraphics[width=0.95\textwidth]{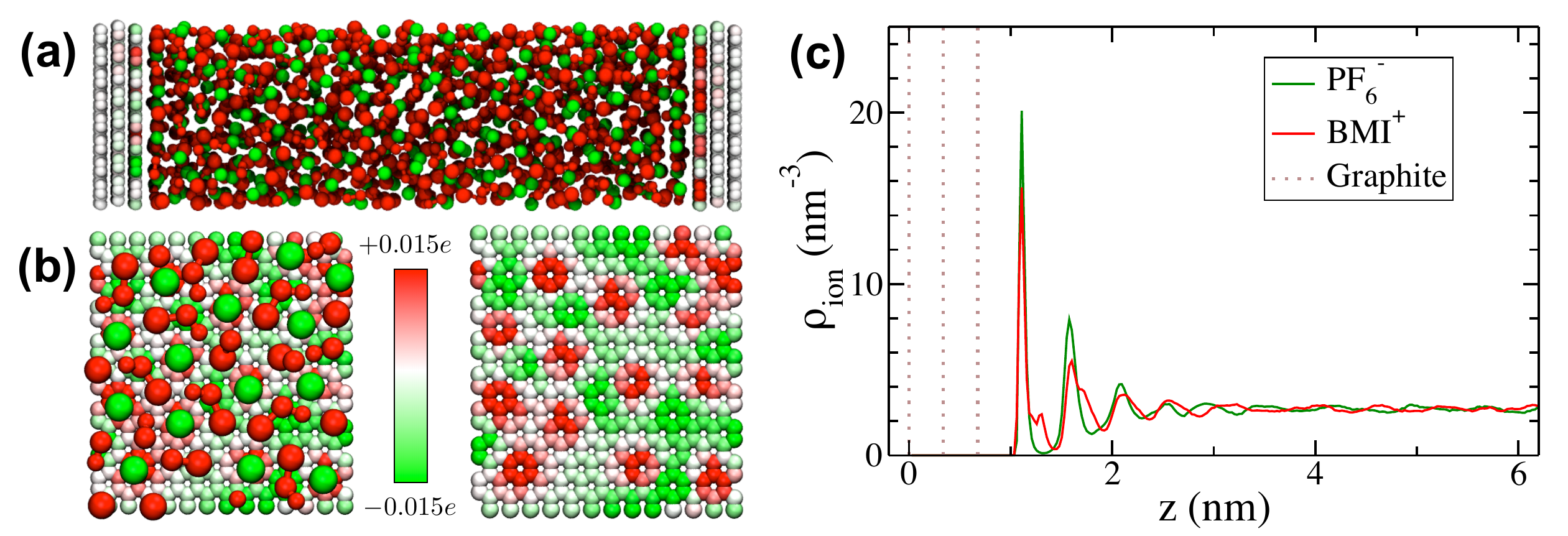}
\end{center}
\caption{
\textbf{Simulated capacitor.}
(a) The capacitor consists of liquid butylmethylimidazolium (red)
hexafluorophosphate (green) and graphite electrodes (white).  
The electrode geometry and force field are detailed in Ref.~\cite{merlet2011a}.
The force field employs the ion-ion interactions from Roy and
Maroncelli~\cite{roy2010a}, and a constant voltage, $\Delta \Psi$, is maintained
throughout a molecular dynamics trajectory by adapting the algorithm of Siepmann
and Sprik~\cite{siepmann1995a}. 
(b)  Polarization of electrode is illustrated
with a configuration of the double layer (left), with molecules drawn smaller
than space filling, and the underlying graphite electrode, with colors
indicative of the instantaneous charge on the carbon atoms.  During
constant-potential simulations, the charge on carbon atoms fluctuates in
response to the thermal motion of the adsorbed liquid. 
(c) The cation and anion density profiles perpendicular to the electrode
surface at $\Delta \Psi$=0V. For the cation the density is plotted for the
center of mass.
}
\label{fig:system}
\end{figure*}

\textbf{Microscopic Models} The specific system we consider is a molecular
simulation model of liquid butylmethylimidazolium hexafluorophosphate
(BMI-PF$_6$) bounded by constant voltage graphite electrodes.  BMI-PF$_6$ is an
example of a room temperature ionic liquid (RTIL) and has been studied
previously~\cite{merlet2011a,merlet2012b,merlet2013b} . 
Figure~\ref{fig:system}(a) shows a characteristic snapshot of
the model electrochemical cell and  Fig.~\ref{fig:system}(b) shows a
characteristic snapshot of the positions of the ions close to the electrode
surface with the corresponding induced charges on the electrode surface, which
fluctuate as a result of the constant potential constraint. The force field we
use to simulate the graphite electrode and the ionic liquid are the same
coarse-grained potentials as used in previous
works~\cite{merlet2011a,merlet2012a,merlet2012b,merlet2013b,roy2010a}.
Similarly, the electrochemical cell is modeled as in previous
works~\cite{merlet2011a,merlet2012b,merlet2013b} with
each electrode consisting of three fixed graphene layers and subject to
two-dimensional periodic boundary conditions~\cite{reed2007a,gingrich2010a}. The
molecular level detail provided by these simulations is evident in
Fig.~\ref{fig:system}(c), where explicit layering within the fluid away from the
electrode arises as a consequence of the discrete nature of the material~\cite{hansen_theory_2006}. 

\textbf{Importance Sampling} In a recent paper, we have introduced a framework
to  analyze molecular simulation data from constant-electrode potential
simulations~\cite{limmer_charge_2013}. We exploit the method of histogram
re-weighting~\cite{ferrenberg_optimized_1989}, which allows us to study the
induced electrode charge as a {\em continuous} function of the applied
potential, rather than by studying it in a series of simulations carried out at
discrete values. The re-weighting method also  enables us to obtain much better
statistical precision than in the previous simulation studies through the
utilization of data from multiple simulations. Such techniques are especially
necessary when studying collective behavior, as timescales for reorganization
make straightforward simulation methods prohibitive. 

To generate configurations for this histogram re-weighing technique, molecular
dynamics simulations were conducted in the NVE ensemble with a time step of 2~fs.
For each simulation, a 200~ps equilibration at $T=400$~K is followed by a 10~ns production run
from which configurations are sampled every 0.2~ps and the corresponding 
total charge $\Qtot=\sigma\times S$ of the electrodes 
(with $\sigma$ the surface charge density and $S$ the surface area)
is used to determine the weight of this
configuration in the ensemble at an arbitrary potential using the weighted
histogram analysis method (WHAM)~\cite{ferrenberg_optimized_1989}.
Application of WHAM allows us to access statistics that would be unobtainable
directly from molecular simulation~\cite{limmer_charge_2013} and to sample the
probability distributions of key variables as continuous functions of the
applied potential.
Simulations were performed for ten potential values 
($\Delta\Psi=0.0$, 0.2, 0.5, 0.75, 1.0, 1.25, 1.5, 1.75, 1.85 and 2.0~V) in
order to ensure a good overlap between the histograms for $\Qtot$, as required
for the histogram re-weighting. 
The distribution of any property $A$ 
is determined as a function of the applied potential from the joint distribution 
of $\Qtot$ and $A$ as 
\begin{align}
P(A|\Delta\Psi)&=\int {\rm d}\Qtot\ P(\Qtot,A|\Delta\Psi) \ .
\end{align}
Specifically, we consider below as observables $A$:
the surface charge density of the total electrode 
and that of subsamples of the electrodes, the numbers of anions and cations
in the first fluid layers adsorbed on both electrodes, the in-plane structure 
factors in these layers and the orientation of cations in these layers.


\section{Results}
\label{sec:results}

\begin{figure}
\begin{center}
\includegraphics[width=0.95\columnwidth]{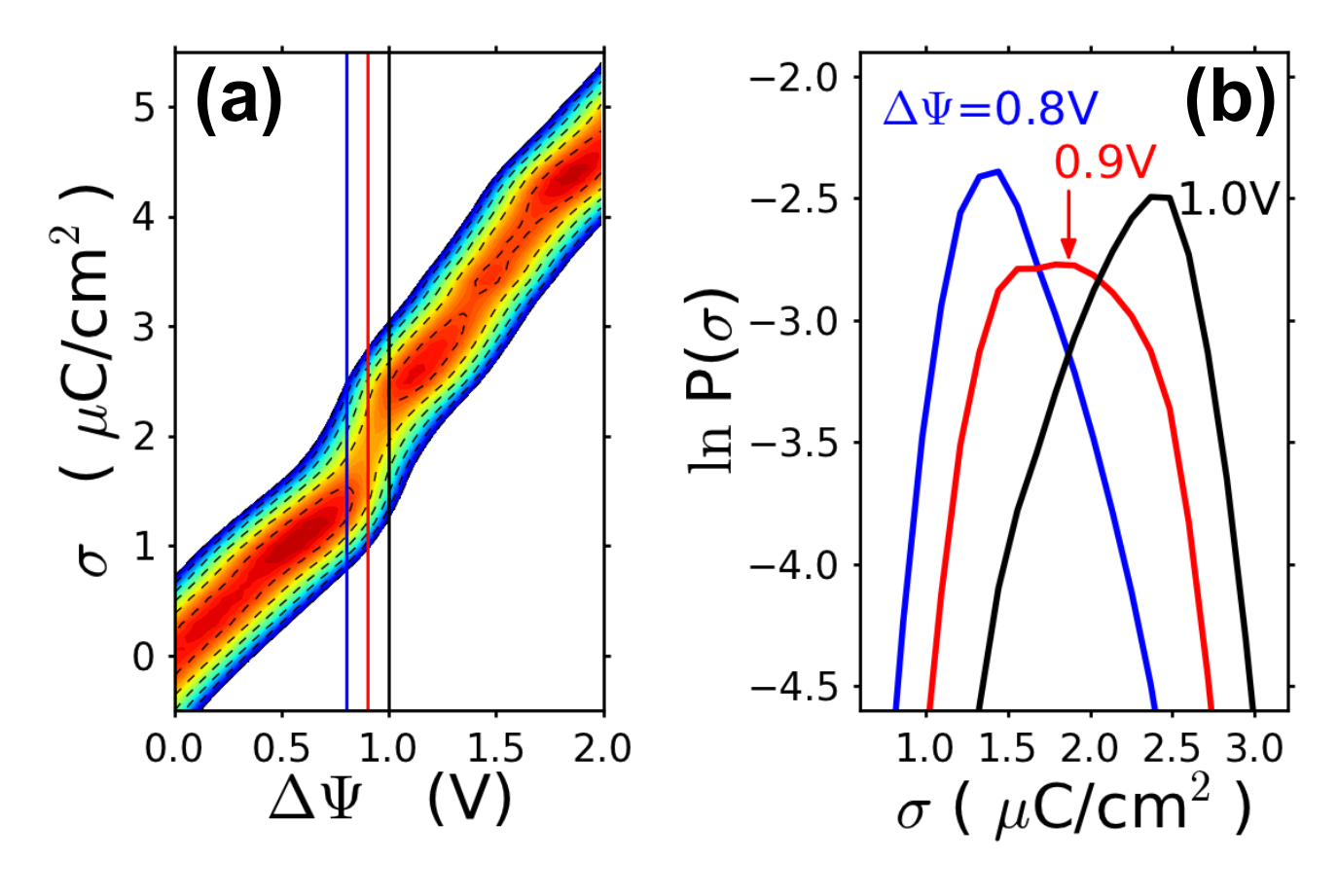}
\end{center}
\caption{
\textbf{Surface charge density distribution.}
(a) The probability
distribution of the charge density $\sigma$ on the electrodes is a function of
the applied potential $\Delta\Psi$. The two-dimensional graph of the
distribution employs a logarithmic scale with lines separated by a difference of
0.5~$k_\mathrm{B}T$
distribution is graphed as a function of $\sigma$ in (b).  Note the fat tails in
the distribution, $P(\sigma)$, and note the markedly non-linear shifts with
changing voltage.}
\label{fig:sigma}
\end{figure}

\textbf{Electrode charge distributions and capacitance} The probability
distribution $P(\sigma | \Delta \Psi)$ of electrode surface
charge density $\sigma$ at various electrode potentials $\Delta \Psi$  is
represented by a contour plot in Fig.~\ref{fig:sigma}a. 
The figure shows the free energy landscape for the system as a function of
surface charge density $\sigma$ and potential $\Delta\Psi$, in
units of Boltzmann's constant, $k_{\rm B}$, times temperature,  $T$. 
It is clear that there are three free energy minima, at $\Delta\Psi \sim$~0.6V, 1.2V, 
and 1.8V, within the voltage range studied, which
correspond to particularly favorable configurations for the interfacial fluid
to balance the corresponding surface charge densities  of
$\sim1.0,~2.6,~4.1~\mu{\rm C/cm^{2}}$. The minima are separated by saddle
points, at 0.9V, for example. A given voltage corresponds to a vertical
slice through the free energy surface and Fig.~\ref{fig:sigma}b shows sections 
for three voltages in the 0.8-1.0~V range.

\begin{figure}
\begin{center}
\includegraphics[width=\columnwidth]{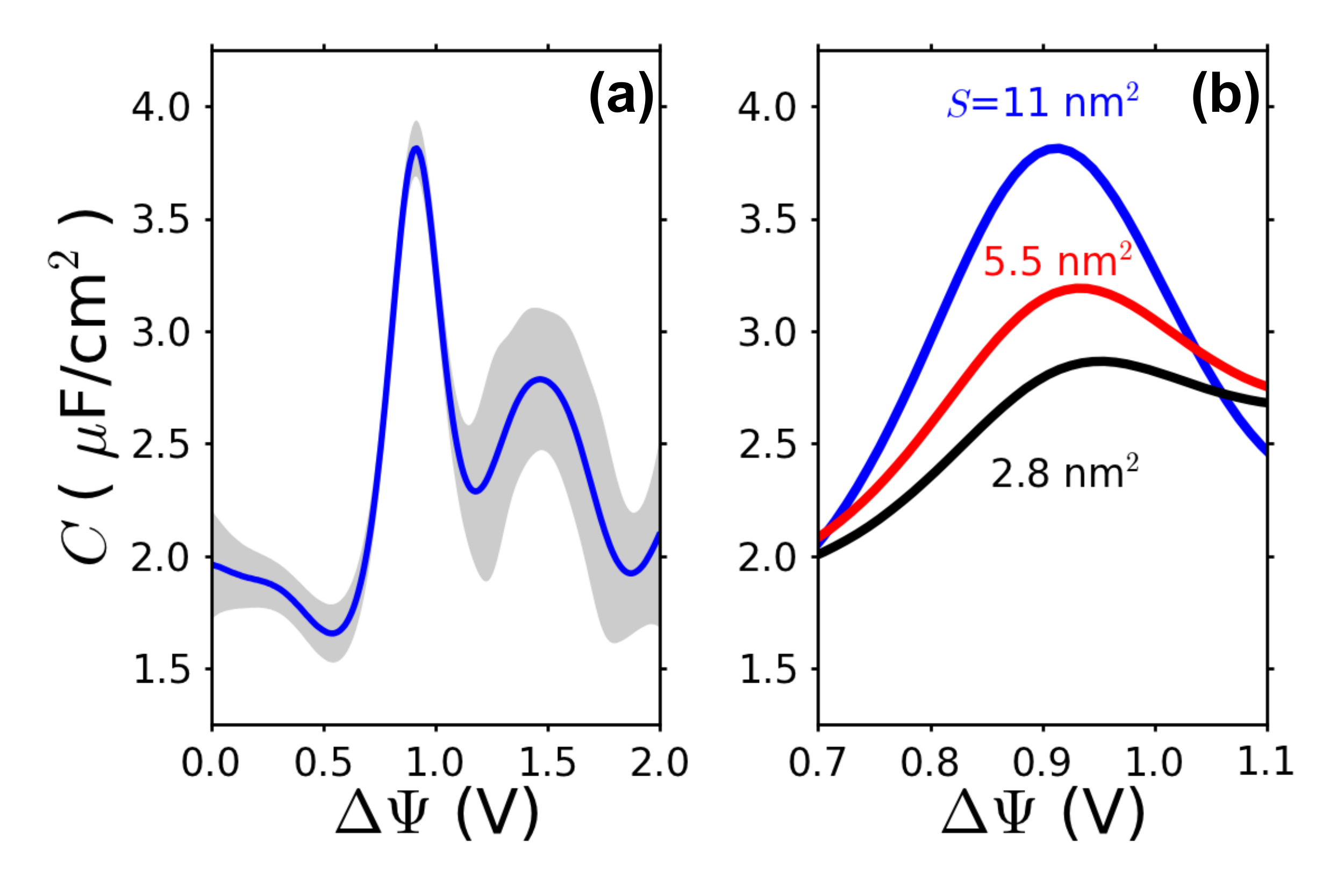}
\end{center}
\caption{
\textbf{Capacitance.}
(a) Differential capacitance, $C$, computed from molecular simulation as a
function of the applied potential $\Delta\Psi$ with an electrode surface area
$S=11$\,nm$^2$. Shaded regions indicate the range of statistical uncertainty.
(b) Growth of computed capacitance peak with growth of electrode surface area.
}
\label{fig:capa}
\end{figure}

The mean-square fluctuations in the electrode surface charge density determine
the differential capacitance $C$ (per unit area) as~\cite{johnson1928a,nyquist1928a},
\begin{equation}
\label{eq:capa-fluct}
C \equiv \frac{\partial \langle \sigma \rangle}{\partial \Delta\Psi } 
= \frac{S}{k_\mathrm{B}T} \,\langle (\delta \sigma)^2 \rangle.
\end{equation}
where $\delta \sigma = \sigma - \langle \sigma \rangle$, and the angle brackets denote
equilibrium average with fixed voltage $\Delta \Psi$.  
\revadd{Note that this relation between the response to voltage and the
equilibrium fluctuations of the surface charge is not limited to the typical
linear response approximation, as both the charge fluctuations and therefore the
capacitance are generally potential dependent~\cite{limmer_charge_2013}.}
As seen in Fig.~\ref{fig:capa}(a), the capacitance exhibits two anomalous peaks,
one near $\Delta \Psi = 0.9$\,V, the other near $\Delta \Psi = 1.5$\,V. Locating
these critical voltages on the free energy surface (Fig.~\ref{fig:sigma}(a) shows that
they correspond to the location of saddle points, where fluctuations
sample two adjacent free energy minima.
Similar features in the differential capacitance have been reported in numerous
experimental
studies~\cite{alam2007a,silva2008a,lockett2008a,roling2012a,druschler2012a,cannes2013a}.

The peaks in capacitance arise from correlations within the
interfacial layer of fluid. This is evident from the electrode surface charge
distributions that exhibit non-Gaussian features
characteristic of a first-order phase transition -- fat tails at conditions away
from phase coexistence and a very flat distribution suggesting incipient 
bi-modality (limited here by a rather small system size) at conditions of
coexistence~\cite{touchette2009large,chandler1987introduction}. 
If the relevant structures in the fluid comprise
correlated domains that are small compared to the net surface area, the distribution
will be Gaussian because the net surface-charge density will reflect
many uncorrelated contributions (the usual Central Limit Theorem argument).
Generally, the distribution, $P(\sigma)$, shifts according to applied
voltage, 
\begin{equation}
\label{eq:distribution}
P(\sigma| \Delta \Psi)\, \propto \, P(\sigma | 0)\,\exp(- \, \sigma \, S \, \Delta \Psi/ k_\mathrm{B}T) \,.
\end{equation}
In the Gaussian case, vertical sections through the
free energy surface, $-k_\mathrm{B}T \ln P(\sigma)$, 
will be parabolas with minima
at the mean value of $\sigma$, and the application of voltage will produce a
proportional shift of that mean - the free energy surface would have the form of
a regular valley with the position of the minimum tracking the voltage smoothly.
This is what has been found previously for simple dielectrics or dilute
solutions~\cite{limmer_charge_2013}.
If the correlated domains
 extend across the entire surface, the distribution will be non-Gaussian, and
the application of voltage will shift the free energy minimum in a markedly non-linear way as different
domain types are stabilized by different voltages.  It is clear that the
surface-charge distributions found in the simulations correspond most closely to
the latter scenario and that, although the non-Gaussian characteristics are a
global property of the free energy surface, they will be most pronounced at
potentials close to 0.9V.

The size of our system is small, so that the non-Gaussian features are
relatively subtle.  Nevertheless, the system is large enough to exhibit size
dependence of response.  In particular, Fig.~\ref{fig:capa}(b) illustrates the
growth of the first $\sim 0.9{\rm V}$ of the anomalous peaks in the differential
capacitance as the observed electrode surface area is increased.  In the absence
of a phase transition, the capacitance $C$ is an intensive property, {\it i.e.},
$\langle(\delta \sigma)^2 \rangle$ scales inversely with $S$.  At conditions of
a coexistence between two phases, however, $C$ will grow with system size
because cooperativeness at a
phase transition extends across the surface.  The growth is significant.
Ultimately, the capacitance at this voltage should increase linearly with $S$,
but the system sizes we have been able to study are not yet large enough to
reach that scaling regime.
The fluctuations we consider might thus foreshadow an unbounded capacitance,
a phenomenon that it not predicted by previous theories even beyond
mean-field~\cite{skinner_capacitance_2010}.

While studying larger system sizes in detail is not possible with
the computer resources currently available to us, we can
nevertheless test the consequences of a putative first order transition.  For
larger systems, one should expect a bimodal distribution
of surface charge density with basins of the charge distribution separated
by a free energy barrier growing with system size. 
We have performed simulations with
a system size increased by 100\% in the $x$ and $y$ directions
for voltages close to the estimated coexistence voltage of 0.9V. For these large
systems, both the ordered and disordered states are metastable over time scales
of 500~ps, and the corresponding charge distributions do not overlap. This
behavior is
consistent with the expected one for a  first order transition.
It raises the usual issue for studying phase transitions in molecular
simulations: proper sampling requiring long simulation times and enhanced
sampling techniques.  
Such an investigation could benefit from
the development of simplified models and analytical theory capable of capturing
the physical phenomena which underpin these effects. 
With these caveats in mind, note that such a transition would also lead to
hysteresis in a macroscopic system, and that hysteresis has been ob- served
experimentally for 
a very similar ionic liquid on graphene using X-ray reflectivity~\cite{uysal_structural_2014}. 

The charge density $\sigma$ is the same on both
electrodes (up to a sign change). Its average value and its variance
(hence the capacitance, see Eq.~\ref{eq:capa-fluct}) reflect the response of
the whole capacitor. 
In order to provide a microscopic interpretation for the observations described
above, we have examined several order parameters based upon averages 
over the ionic positions at each electrode. Given the difference in shape and
size between cations and anions, 
the microscopic structure of the interface is likely to differ at positive and
negatively charged electrodes.  In the following, results are reported as a
function of 
the variable $\Psi$ defined as $+\Delta\Psi/2$ (resp. $-\Delta\Psi/2$)
for the positive (resp. negative) electrode when the capacitor is
submitted to a voltage $\Delta\Psi$. Since in the present system the 
point of zero charge is close to 0~V, \revadd{with comparable capacitances
for both interfaces~\cite{merlet2011a}, $\Psi$ approximately} 
reflects the actual potential drop across each double layer. 

\textbf{Out-of-plane layering} Figure \ref{fig:system}(c) illustrates the density profiles
for anions and cations at zero applied potential. Examples of the density
profiles at higher potentials for this system have been shown elsewhere
~\cite{merlet2013a} and layering similar to that indicated in the figure
persists over the full range of potentials studied here. Although the layering
is a generic property of dense fluids, the arrangement of the ions
within the layers depends on the details of the interionic interactions. In the
present case, at 0V, the maxima in the principal peaks of the anion and cation
densities coincide, which shows that both ions have roughly the same size and
may approach the surface equally closely (the small shoulder appearing on the
cation peaks arises from different possible orientations of the cations on the
surface). If the ions were of very different size, the cation and anion layers
could be displaced from each other, as seen \textit{e.g.} in Ref.~\cite{feng2011b}. 
Furthermore, from the areas of corresponding cation and
anion peaks for the first two layers it follows that a given layer contains a
very similar number of cations and anions (a feature we explore in detail
below). This will be favored (at low electrode surface charge, and in the
absence of strong attractive interactions of one species with the surface) by
the coulombic interactions between the ions which will try to impose local
charge neutrality. 

\begin{figure*}[ht!]
\begin{center}
 \includegraphics[width=.85\textwidth]{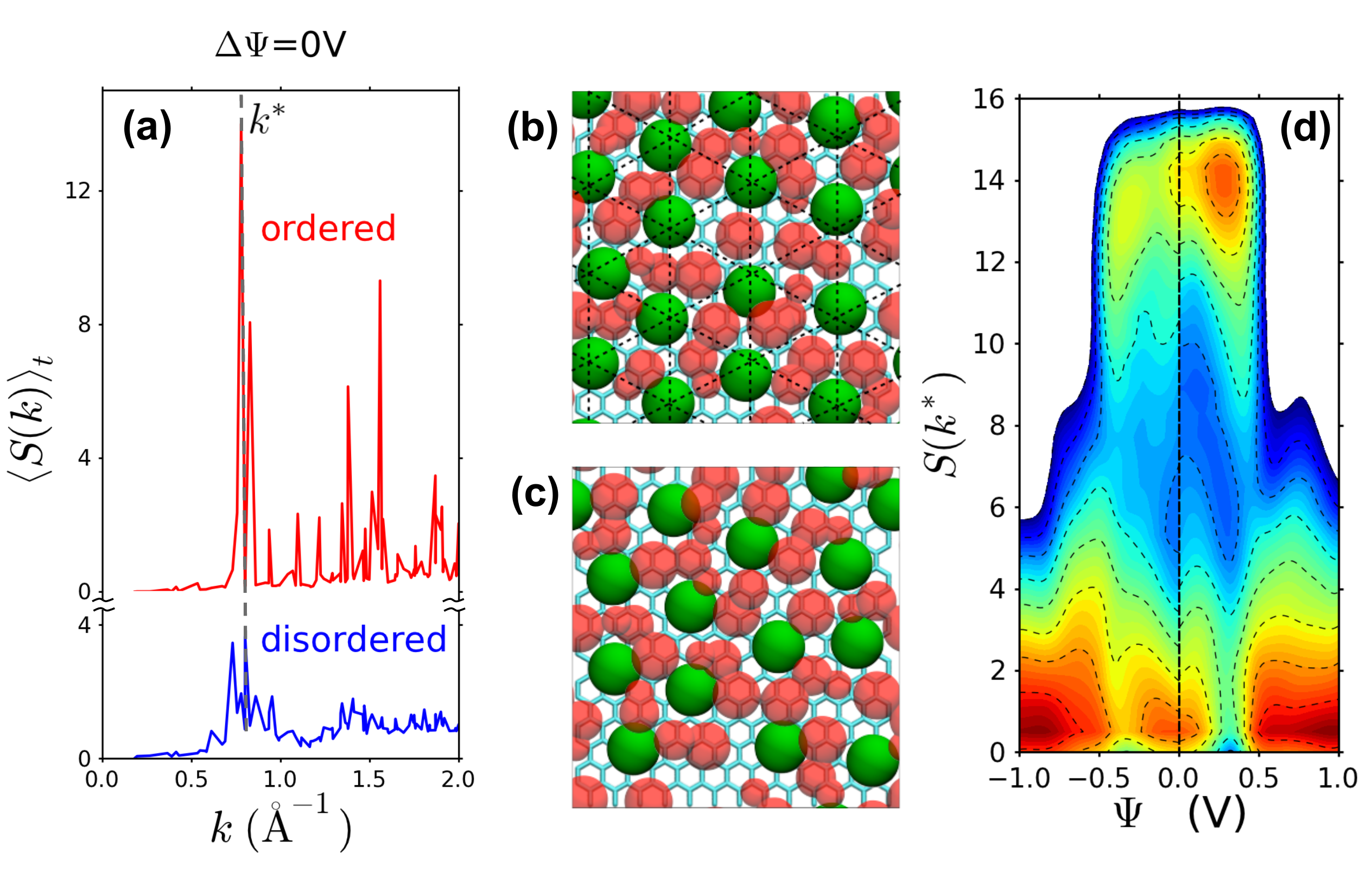}
\end{center}
\caption{
\textbf{In-plane structure of the adsorbed layer}
(a) Average anion-anion structure factor $\langle S(k)\rangle$ in the first
adsorbed layer on the electrodes, as a function of the norm of the wave vector,
for a voltage $\Delta\Psi=0$V. Structure factors for ordered and disordered
states, which are both observed at this particular voltage, are averaged over 1 ns.
(b) Configuration illustrating the adsorbed layer in the ordered state.  
(c) Configuration of a disordered state.  
(d) Probability distribution of the anion-anion structure
factor at the maximum $S(k^*)$ in the first adsorbed layer on the
electrodes, as a function of 
$\Psi$, defined as $+\Delta\Psi/2$ (resp. $-\Delta\Psi/2$)
for the positive (resp. negative) electrode when the capacitor is
submitted to a voltage $\Delta\Psi$.
The probability is reported
on a logarithmic scale, with lines separated by a difference of 0.5.  
Note the presence, on both electrodes, of three distinct basins, indicative of
i. a low-voltage phase where the double layer is disordered, ii. an
intermediate-voltage phase where the double layer is ordered, and iii. a
high-voltage disordered phase where the double layer is charge unbalanced. 
}
\label{fig:skmax}
\end{figure*}

\textbf{In-plane order} The in-plane anion structure factor for the anions 
in the first adsorbed layer on a given electrode is
\begin{equation}
S({\bf k})= \frac {1}{N_-}\sum_{i,j} e^{i{\bf k\cdot r}_{ij}} \;, 
\end{equation}
where the sum runs over pairs of anions that are found in the first layer of
fluid adjacent to the electrode surface and ${\bf k}$ is a wavevector which lies
in the plane of the electrode and is commensurate with the periodic boundary
conditions. These wavevectors form a discrete lattice and in 
Fig.~\ref{fig:skmax}a we have plotted the average value $\langle S\rangle$ 
as a function of $|{\bf k}|$
for two states visited along a the trajectory at 
$\Delta\Psi=0$V. At a given voltage, the interface may explore configurations
with varying degrees of in-plane order. The bottom part corresponds to a fluid-like
configuration, whereas the top one shows a strong, Bragg-like peak at
$k^*\simeq 0.8~{\rm \AA}^{-1}$, which corresponds to $2\pi$ divided by the mean
nearest-neighbor separation of the anions in the first layer, and strongly
suggests a 2-d lattice-like organization for the anions.
 
Figures~\ref{fig:skmax}b and~\ref{fig:skmax}c show snapshots which illustrate
such ordered and disordered structures, respectively. The relative contribution
of these structures to an averaged property of the interfacial fluid, like the
in-plane structure factor, depends on the voltage, as we shall see below. In
particular,
the lattice-like ordering only persists for a certain range of applied potentials.

The length scales of the underlying graphite play no role
in this ordered structure.  Indeed, structures like it are inferred from experiments of
monolayers of  BMI-PF$_6$ on gold~\cite{pan2006a}.  
This is not to say that other systems cannot show an effect of commensurability
between in-plane order in the electrolyte and the
surface~\revadd{\cite{kornyshev_phase_1995,tazi2010a}}.  
But the role of graphite in the system studied here is only that of a polarizable metal.  
Note also that the geometry of the triangular lattice is distinct from that of the (nearly)
square electrode that is periodically replicated in the simulation.  Thus, the
ordered state we find in the simulation is not a finite-size artifact of the
simulation geometry.

In Fig.~\ref{fig:skmax}d we show a contour plot of the probability
distribution of values of $S(k^*)$  as a function of applied potential. This is
again plotted as -$k_{\rm B}T\ln P$ so that the surface can be interpreted as a
free energy landscape in the structural order parameter $S(k^*)$. 
This plot reveals a substantial basin of stability for the ordered
state of the first interfacial layer of fluid for potential differences 
$\Delta\Psi$ between about 0.5 and 0.75V (remember that
$\Psi=\pm\Delta\Psi/2$). This basin is present on both electrodes,
even though the ordered state is less probable on the negative electrode. 
At both higher and lower voltages the fluid layer is translationally
disordered. This observation strongly suggests that the peak in the differential
capacitance observed as the potential is increased through 0.9V is associated
with the transition between an ordered, lattice-like state of the first layer
and the disordered state. The first-order nature of this phase transition,
which is evident from free energy surface in $S(k^*)$, shows up in the statistics
of the electrode surface-charge fluctuations and the resulting peak in the
differential capacitance.
However, the transition from the low potential,
disordered state to the ordered state, which might be expected at about 
$\Delta\Psi=0.35$V,
does not appear to affect the surface charge significantly. Although a
transition in first-layer structure is seen in $S(k)$, it does not appear to be
strongly coupled to the surface charge. 

In the present case, the transitions on both electrodes occur approximately
at the same voltage $\Delta\Psi$. For systems where the difference in size and shape between
cations and anions is stronger, or where the fine structure of the electrode
plays a more important role, such transitions may occur for different voltages
on each electrode. For example, previous work on electrochemical cells
involving a molten salt, LiCl, and (100) aluminum electrodes, an order-disorder
transition occurred in the first adsorbed layer close to the potential of zero
charge\revadd{~\cite{tazi2010a}}. For positive surface charges, an ordered structure, commensurate to the
underlying metal structure was stabilized while the structure was liquid-like
otherwise. As a result, a large step was observed on the charge vs. potential
plot, which should be in principle associated to a large peak in the
differential capacitance. Such a peak could not be evidenced since the
importance sampling tools discussed here had not been introduced at that time.
The asymmetric behavior could be attributed \revadd{in that case} 
to the large difference in
polarizability between the lithium and chloride ions. As soon as the
polarization effects (on the ions) were turned off, the ordered structure was
unable to form, leading to an almost constant differential capacitance over the
whole simulated potential range~\cite{tazi2010a}.

\begin{figure}[t]
\begin{center}
\vspace{0.5cm}
\includegraphics[width=0.95\columnwidth]{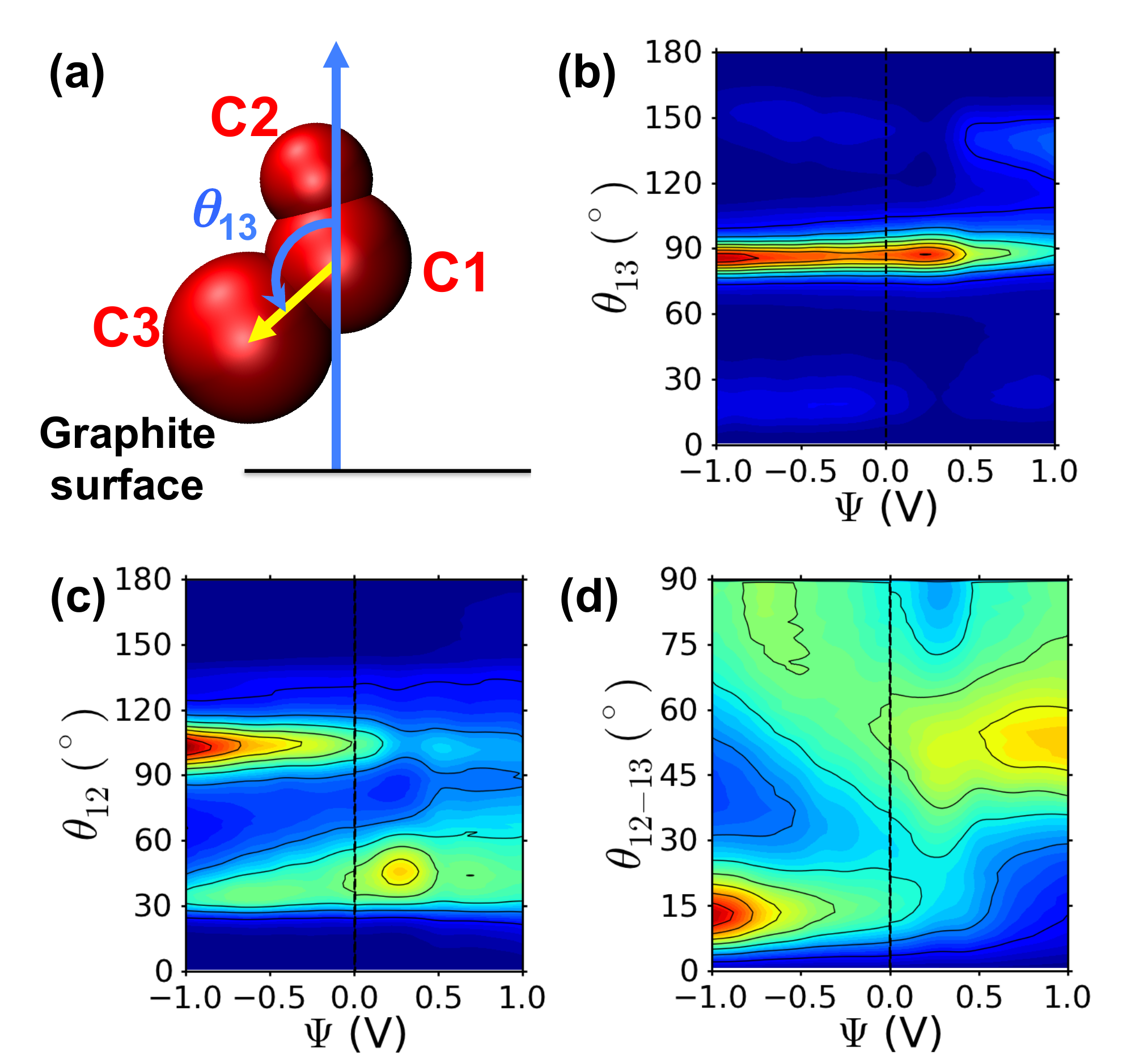}
\vspace{-.5cm}
\end{center}
\caption{\label{fig:angle}
\textbf{Orientation of cations on the electrodes.}
(a) Definition of the angle $\theta_{13}$ 
between the imidazolium-butyl axis and the normal to the graphite 
surface (into the fluid). 
The space-filling representation of the cation illustrates
the three sites defined in the force field to model the imidazolium ring
(C1) as well as the butyl (C3) and methyl (C2) side chains.
The $\theta_{12}$ angle is defined similarly, while the 
$\theta_{12-13}$ angle corresponds to the angle with the normal to the molecular
plane.
(b) Probability distribution of the angle $\theta_{13}$
as a function of $\Psi$, defined as $+\Delta\Psi/2$ (resp. $-\Delta\Psi/2$)
for the positive (resp. negative) electrode when the capacitor is
submitted to a voltage $\Delta\Psi$.
The probability is reported
on a logarithmic scale, with lines separated by a difference of 0.5.
(c) and (d) Same as (b) for the angle $\theta_{12}$ and $\theta_{12-13}$.
}
\end{figure}

\textbf{Cation orientation and stoichiometry of the layers} The orientation of
cations in the first ionic layer on the electrode 
evolves with the applied potential $\Delta\Psi$. This can be measured
using angles such as the one 
between the imidazolium ring-butyl (C1-C3) axis and the normal to the surface,
illustrated in Fig.~\ref{fig:angle}(a). The distributions of these angles
as a function of $\Psi$ are reported in Figs.~\ref{fig:angle}(b),
\ref{fig:angle}(c) and~\ref{fig:angle}(d).
On both electrodes, the C1-C3 axis remains almost parallel to the surface
for all voltages. On the contrary, the distribution of the ring-methyl 
(C1-C2) orientation displays bimodality, with relative weights of the two
configurations that evolve with the applied voltage. The cation may
either lie flat on the surface ($\theta_{12}\sim90^\circ$) or tilt with the methyl group
pointing outward ($\theta_{12}\sim30^\circ$) while the imidazolium ring and butyl
group remain on the surface ($\theta_{13}\sim90^\circ$). At 0~V, both orientations
have similar probabilities. With increasing voltage, the fraction of tilted
cations increases on the positive electrode and decreases on the negative one.
This behavior is consistent with the electrostatic interaction between the positive
partial charge of the methyl group and the electrodes. The fact that the bulkier
butyl and imidazolium ring remain closer to the surface may be due to packing
effects, their smaller charge density compared to the methyl group
and to the stronger attractive dispersion interaction with the wall,
\revadd{in agreement with observations for a model cation consisting of a
charged head and a neutral tail~\cite{georgi_anatomy_2010}.}

\begin{figure}
\begin{center}
 \includegraphics[width=\columnwidth]{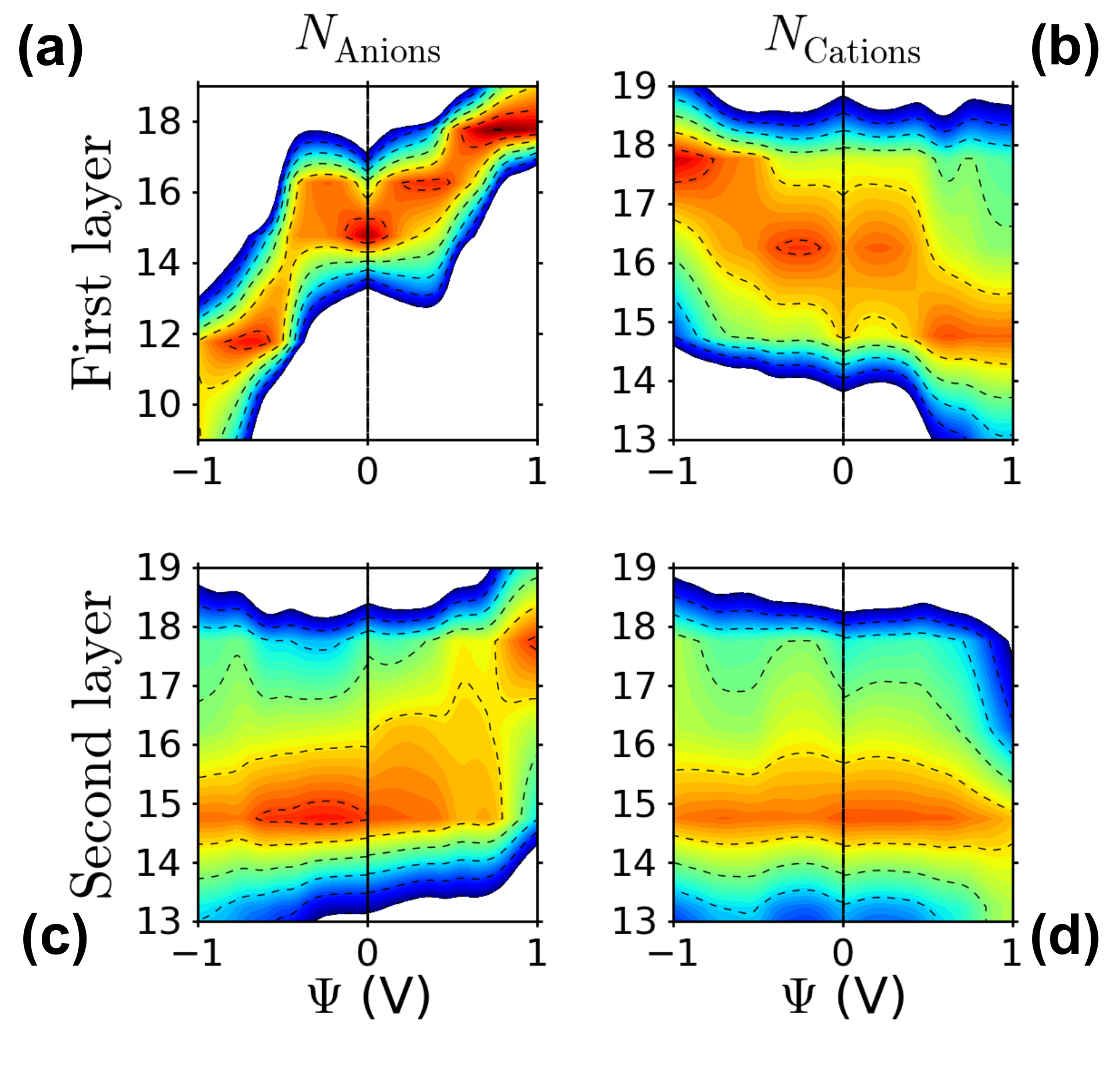}
\end{center}
\caption{
\textbf{Number of ions in the adsorbed layers}
Probability distribution of the number of anions $N_{\rm Anions}=N_-$
(a,c) and
cations $N_{\rm Cations}=N_+$ (b,d) in the first (a,b) and second (c,d)
fluid layers on both electrodes, as a function of 
$\Psi$, defined as $+\Delta\Psi/2$ (resp. $-\Delta\Psi/2$)
for the positive (resp. negative) electrode when the capacitor is
submitted to a voltage $\Delta\Psi$.
The distribution is reported on a logarithmic scale, 
with lines separated by a difference of 0.5.
}
\label{fig:nions}
\end{figure}

The value of the electrode surface charge at a given potential reflects the
spatial distribution of the induced charge density in the interfacial layer of
electrolyte. In order to relate this to the structural changes in the fluid, we have
plotted in Fig.~\ref{fig:nions} the probability distributions for finding a
given number of anions ($N_-$) and cations ($N_+$) in the first and second
layers of fluid. This is plotted
as  $-k_\mathrm{B}T \ln P(N)$ so that it can be interpreted as a free-energy
landscape.
The most striking features of these plots are the pronounced changes in the most
probable number of anions in the first layer at both anode and cathode that
occur as the voltage is increased. The preferred numbers of both species in the
{\emph second} layer remains roughly constant, at least until the overall cell voltage ($\Delta\Psi$)
exceeds 1.3V - corresponding to $\Psi \sim 0.65$V.  
The preferred cation number in the first layer remains constant
until $\Delta\Psi$  reaches 1.0V and there is initially, close to zero volts, an apparent
charge imbalance in the first layer because the number of cations exceeds that
of anions.
 
When the voltage $\Delta\Psi$ exceeds about 0.35V the number of anions at both electrodes increases
so that it balances the number of cations. This occurs because cations become increasingly 
orientationally ordered by larger applied
potentials and this allows enough space for the additional anions to enter the
first layer. As the ions become tightly packed in the first
layer, the structure factor develops its Bragg-like peak (Fig.~
\ref{fig:skmax}) showing that the lattice-like arrangement is associated with
the ideal stoichiometry.  Note that the capacitance (Fig.~\ref{fig:capa}) is a
decreasing function of  potential between zero and about 0.6V, suggesting that
the interfacial fluid becomes increasingly resistant to charging over this
range. At $\Psi$ about 0.5V, there is a sharp increase, over a narrow voltage window,
in the number of anions at the positive electrode, which is partially
compensated by a smaller decrease in the number of cations. At the cathode,
there is a more extended transition at about $\Psi \sim -0.5$V 
which results in still larger changes in the
numbers of both anions and cations.  These rapid increases in the magnitude of
the  interfacial charge close to the electrode surface are  responsible for the
peak in the capacitance close to an overall cell voltage drop of $\Delta\Psi$ =0.9V. 
 
From Figs.~\ref{fig:skmax} and~\ref{fig:nions} it is
clear that the nature of the structural transitions in the first layer that are
responsible for the capacitance peak at 0.9V are collective transitions
between stoichiometric ($N_+=N_-$), lattice-like ordered states and
non-stoichiometric, or partially demixed, disordered states with the latter
persisting to higher potentials. A possible interpretation is that the
increasing magnitude of the potential drives the like-charged ions out of the
first-layer and causes the melting to the non-stoichiometric disordered phase.
Notice that the capacitance peak is associated with the {\it transition} out
of the ordered state, rather than the existence of the ordered state itself:
from the free energy landscape in Fig. \ref{fig:skmax} it is clear that the
lattice structure is most stable for a voltage $\Delta\Psi$ of about 0.6-0.75V,
substantially lower than that at which the capacitance peak is observed.

At higher potentials, in particular near to the second peak in the capacitance
at $\Delta\Psi \sim$1.45V (Fig.~\ref{fig:capa}), the structure factor 
(Fig.~\ref{fig:skmax})  suggests that the interfacial fluid layers are laterally
disordered. The ion numbers in the first adsorbed layer on both electrodes are
almost unchanged close to this voltage (Fig.~\ref{fig:nions}). Only in the
second adsorbed layer can changes be observed, in particular for the number of
anions on the positive electrode,
but it is difficult to draw definitive conclusions from our data on the
microscopic mechanisms underlying the second peak in capacitance. 

\section{Conclusion}

The present work illustrates that the interface between an ionic liquid
and a metal electrode can exhibit structures and fluctuations that are not
simple reflections of surrounding bulk materials. This rich behavior is absent
from a mean-field picture that averages over intra-planar structure. 
The charge of the electrode is screened by the interfacial fluid and induces subtle 
changes in its structure. In the present case with ions of comparable sizes
but different shapes, small voltages induce changes in the orientation of
the anisotropic cations which differ on the two electrodes. In turn, the effective
lateral density changes, allowing for the in-plane ordering of the anions and
cations in the first adsorbed layer. This in plane order is analogous to that
found in experiment~\cite{liu_coexistence_2006}. The transition out of this ordered state
into a disordered non-stoichiometric state is associated with a large change in
the charge distribution across the interfacial region of the electrolyte and
leads to a peak in capacitance. While a similar peak in the capacitance had been
also noted in experiment~\cite{su_double_2009}, its origin had been
debated~\cite{lockett2010b,druschler2012a}. This work offers a microscopic
explanation, free from the experimental difficulties such as surface
reconstruction and defects that plague such experimental studies. Moreover, the
presence of ordered states like those we describe here may also be at the origin
of the slow capacitive process that have been observed when recording broadband
capacitance spectra~\cite{roling2012a,druschler2012a}.  The structural changes
are likely to also induce different mechanical responses. The approach developed
in the present work should thus also provide insights into microscopic mechanism
for  voltage-dependent lubricants~\cite{sweeney_control_2012} and RTIL-based
actuators~\cite{liu_actuator_2013}.


\section*{Acknowledgements}
BR and DC acknowledge financial support from the France-Berkeley Fund
under grant 2012-0007.  DTL was supported in the Helios/SERC project by the
Director, Office of Science, Office of Basic Energy Sciences, and by the
Division of Chemical Sciences, Geosciences, and Biosciences of the U.S.
Department of Energy at LBNL at Lawrence Berkeley National Laboratory under
Contract No. DE-AC02-05CH11231.
CM, MS and BR acknowledge financial support from the French Agence Nationale de
la Recherche (ANR) under grant ANR-2010-BLAN-0933-02. 
We are grateful for the computing resources on JADE (CINES, French National HPC)
obtained through the project x2012096728.



\end{document}